\documentstyle[prb,aps,psfig,multicol]{revtex}

\begin{document} 
\preprint{subm. to 44rd Conference on Magnetism and Magnetic Materials, Nov. 15-18, 1999, San Jose} 
\draft 

\title{Hall effect of epitaxial double-perovskite Sr$_{2}$FeMoO$_6$ thin films}

\author{W.~Westerburg, F.~Martin, and G.~Jakob}  

\address{Institute of Physics, University of Mainz, 55099 Mainz, Germany} 

\date{29 July 1999} 

\maketitle 

\begin{abstract} 
We prepared high epitaxial thin films of the compound Sr$_{2}$FeMoO$_6$ 
with narrow rocking curves by pulsed laser deposition.
The diagonal and nondiagonal elements of the resistivity tensor were 
investigated at temperatures from 4\ K up to room temperature in magnetic 
fields up to 8\ T.
An electronlike ordinary Hall effect and a holelike anomalous Hall 
contribution are observed. Both coefficients have reversed sign 
compared to the colossal magnetoresistive manganites. 
We found at 300\ K an ordinary Hall coefficent of $-1.87\times10^{-10}$\ m$^3$/As, 
corresponding to a nominal charge carrier density of four 
electrons per formula unit. 
At low temperature only a small negative magnetoresistance is observed which 
vanishes at higher temperatures.
The temperature coefficient of the resistivity 
is negative over the whole temperature range. A Kondo like behavior is observed 
below 30 K while above 100 K variable range hopping like transport occurs.
\end{abstract}  



\begin{multicols}{2}
\section{ Introduction}
Half-metallic ferromagnetic oxides as the manganites have reattracted much interest.
Recently, a large negative magnetoresistance (MR) in a further oxide Sr$_{2}$FeMoO$_6$ (SFMO) was observed 
\cite{Kobayashi98}. This compound has an ordered
double-perovskite structure and is, like the manganites, a ferrimagnetic (or ferromagnetic \cite{garcia99}) oxide with
a Curie-temperature of 410-450\ K and a highly spin-polarized conduction band. For applications 
as a magnetic field sensor at room temperature a high spin-polarization is necessary to obtain a large magnetoresistance
in low fields \cite{wester99}. Therefore spin-polarized magnetic compounds with Curie temperatures 
well above 300\ K are interesting. 
We prepared epitaxial thin films of SFMO and investigated their structural and magnetic
properties. Further, the longitudinal and transverse resistivity were measured as a function of temperature and
magnetic field.\\ 
\section{ Experiment}
\begin{figure}[t]
\psfig{file=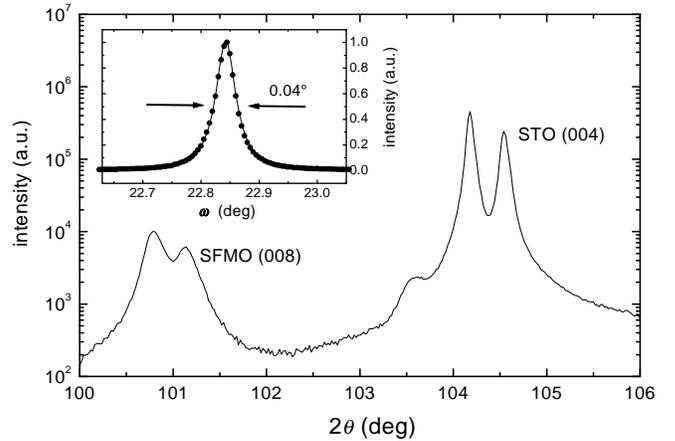,width=1.0\columnwidth} 
\vspace{0.5em} 
\caption{$\theta/2\theta$-scan of the SFMO thin film showing the (008) reflection peak
         (Cu $K_\alpha$ doublet). The three peaks at higher angles are due to the substrate. 
         In the inset a rocking curve (2$\theta=45.32^\circ$) with an extremely narrow 
         linewidth ($\Delta{}\omega{}=0.04^{\circ}$),
         indicating a high degree of epitaxy, is shown.}
\label{scan} 
\end{figure}

\begin{figure}[t]
\psfig{file=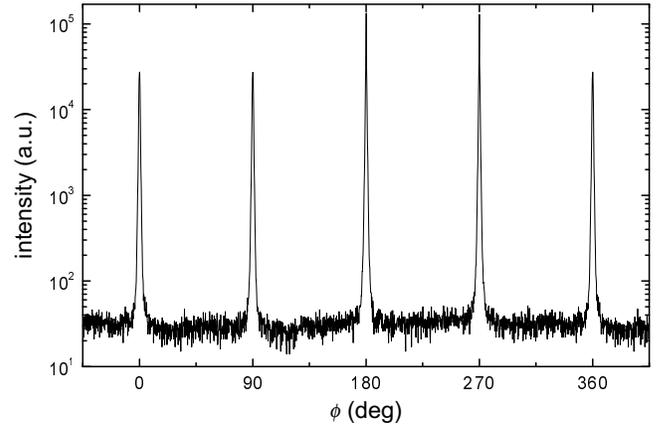,width=1.0\columnwidth} 
\vspace{0.5em} 
\caption{$\phi$-scan of the SFMO\{224\} reflections of the thin film grown on a SrTiO$_3$ substrate. 
        The 4-fold in-plane symmetry, the film axes are parallel to
        the substrate axes, is visible.}
\label{phi} 
\end{figure}

Using pulsed laser deposition we prepared epitaxial thin films of SFMO 
from a stoichiometric target on (100) SrTiO$_3$ (STO) substrates. 
The substrate temperature during deposition was 700$^{\circ}$C in an oxygen
partial pressure of $10^{-3}$ Pa. In x-ray diffraction only film 
reflections corresponding to a (00$\ell$) orientation are visible.
In Fig.\ \ref{scan} a detail of the
$\theta/2\theta$-scan, showing the (008) reflection peak of SFMO nearby 
the (004) reflection peak of STO, is presented.
A high degree of orientation of the $c$-axis is achieved. 
Rocking angle analysis of the SFMO (004) reflection shows an angular spread of 0.04$^{\circ}$, as
can be seen in the inset of Fig.\ \ref{scan}. 
The in-plane orientation was investigated by $\phi$-scans using the 
\{224\} reflections. The film axes are parallel to the substrate axes with
a 4-fold in-plane symmetry, demonstrated in Fig.\ \ref{phi}.
With this preparation conditions the $c$-axis is elongated to 7.998\ \AA{} 
compared to bulk material \cite{Kobayashi98}. 
This indicates either epitaxial strain between SFMO and the STO 
substrate or (and) a non stoichiometric oxygen content of the films.\\ 
With scanning electron microscopy and atomic force microscopy we 
found a smooth surface with a roughness of 10\ nm and the average grain 
size to be 50\ nm. By Rutherford backscattering on a reference sample on MgO
substrate the metal atom stoichiometry was 
determined to be Sr$_{2}$Fe$_{1.06}$Mo$_{0.90}$O$_x$. This 
is within the experimental errors identical to the nominal composition of the
target (Sr$_{2}$FeMoO$_{6\pm\delta}$). The film thickness was 250\ nm, measured with 
a Mireau interferometer. The samples, not resistant against water, 
were patterned photolithographically and etched to a Hall bar structure. 
The longitudinal and transverse resistivity were measured in magnetic 
fields up to 8\ T from liquid helium temperature to 300\ K. 
A standard four point technique with DC current was used.
The Hall coefficient was determined by slowly sweeping the magnetic field 
in positive and negative field direction with an asymmetric current 
injection to minimize the parasitic longitudinal voltage on the Hall 
contacts \cite{Fritzsche59,Jakob98}. Spontaneous magnetization was determined in small fields 
($B=100$\ mT) with a SQUID magnetometer.

\section{ Results and Discussion}
\begin{figure}[t]
\psfig{file=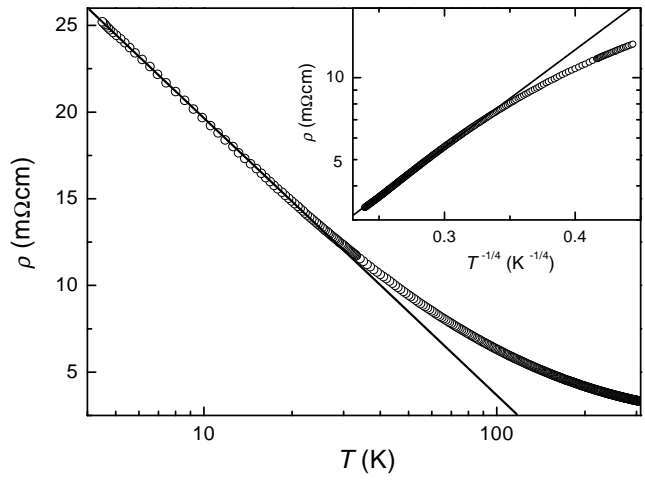,width=1.0\columnwidth} 
\vspace{0.5em} 
\caption{Temperature dependence of the resistivity $\rho_{xx}$ in zero field. 
	 Below 30 K the resistivity increases logarithmic
         with falling temperature. The variable range hopping
	 model in the inset describes the data above 100 K. The lines are in
	 both cases a guide to the eye.}
\label{RvT} 
\end{figure}

The zero field resistivity is presented in Fig.\ \ref{RvT}. 
The room temperature resistivity value of 3 m$\Omega$cm is comparable 
with other reported results \cite{Manako99}, but down to 4 K the resistivity increases 
almost an order of magnitude. Depending on preparation the sign of the 
temperature coefficient and the sign of the MR
changes \cite{Asano99}. We observed in our semiconducting films a 
negative MR of -3.5 \% at a temperature of 4 K and a magnetic 
field of 8 T, similar to the result of Asano {\it et al.} \cite{Asano99}. 
At higher temperatures the MR vanishes. While in 
the plot of Fig. \ref{RvT} the resistivity shows a smooth temperature 
variation with a negative temperature coefficient, closer inspection 
indicates a change in conduction mechanism.
Below 30 K the resistivity increases strictly logarithmic
with falling temperature. This behavior is known from Kondo systems,
where it results from carrier scattering at uncorrelated magnetic 
impurities. Above 100 K the temperature dependence of the resistivity is 
best described by $\rho \propto \exp((T_0/T)^{0.25})$ with $T_0=4900$\ K, as
can be seen in the inset of Fig.\ \ref{RvT}.
The origin of the opposite behavior in transport properties between different epitaxial
films remains up to now unclear. One cause may be order-disorder effects in the B-cation
arrangement of the A$_2$B'B''O$_6$ double-perovskite. A segregation into clusters of
compositions SrMoO$_3$ ($a=3.975$ \AA )\cite{Brixner60} and SrFeO$_3$ ($a=3.869$ \AA )\cite{Yakel55} 
should be visible in x-ray diffraction as severe peak broadening for small clusters 
or as peak splitting for large clusters. Both effects are not observed. The peak splitting 
visible in Fig.\ \ref{scan} for substrate and film peak is due to the Cu $K_\alpha$ doublet only. 
For the double-perovskites random, rock salt and layered B',B'' sublattice types are known 
\cite{Anderson93}. 
In neutron powder diffraction of SFMO Rietveld refinement showed perfect B',B''-rock salt
structure \cite{garcia99}.
Due to extinction conditions the x-ray diffraction in Bragg-Brentano geometry cannot
resolve the sublattice structure for (00$\ell$) oriented films.  
However, the saturation magnetization of our films is with one $\mu_{\rm B}$ per
formula unit (f.u.) 
much smaller than the moment of 3 $\mu_{\rm B}$/f.u. observed by Kobayashi {\it et al.} \cite{Kobayashi98} 
in a bulk sample.
This discrepancy is a hint to cation site disorder \cite{Ogale99}.   
Other possible influences as non stoichiometric oxygen content and substrate induced strain
will be the subject of future investigations.\\

\begin{figure}[t]
\psfig{file=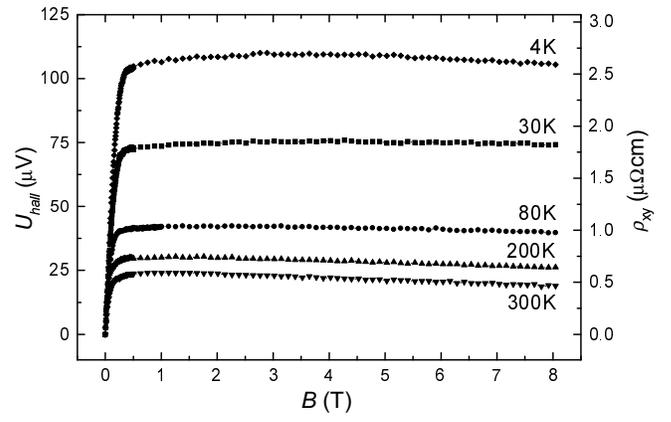,width=1.0\columnwidth} 
\vspace{0.5em} 
\caption{Hall voltages $U_{hall}$ (left axis) and Hall resistivities $\rho_{xy}$ (right axis) as functions of magnetic 
	field for several constant temperatures.}
\label{Hall} 
\end{figure}

In ferromagnetic materials the transverse resistivity is given by
\begin{equation}
	\label{RhoHall}
	\rho_{xy}=R_H B+R_A\mu_0 M
\end{equation}
with the magnetization $M$ and the ordinary and anomalous Hall coefficients $R_H$ and $R_A$, 
respectively \cite{Karplus54}.
The Hall voltage $U_{hall}$ was measured at several constant temperatures between 
4 K and 300 K in magnetic fields up to 8 T. 
Fig.\ \ref{Hall} shows the results for the Hall resistivity and the Hall voltage. 
The error in the data is smaller than the symbol size. The low magnetoresistivity 
of both the sample and the Pt-thermometer allows at high temperatures the elimination of 
the parasitic longitudinal part of the Hall voltage and a quantitative analysis of $R_H$. 
Due to the increasing resistance the measurement current 
was reduced from 1 mA to 100 $\mu$A for the data taken at $T=4$ K, leading to a worse signal to noise ratio. 
In low fields, a steep increase of $U_{hall}$ with increasing field is seen. This part, where the
magnetization of the sample changes, is dominated by the anomalous Hall contribution. 
In the case of SFMO $R_A$ is holelike, in contrast to the manganites \cite{Jakob98}.  
At 1 T a maximum occurs and at higher fields the data show a linear negative slope. 
In this high field regime the
magnetization of the sample is constant and therefore, according to Eq.\ \ref{RhoHall}, 
the ordinary Hall effect becomes visible. This behavior, positive $R_A$ and negative $R_H$, was also
observed in iron and ferromagnetic iron alloys \cite{Bergmann79}.
The linear negative slope
${\rm d}\rho_{xy}/{\rm d}B$ indicates an electronlike charge-carrier concentration.
The Hall coefficent at 300 K is $-1.87\times10^{-10}$\ m$^3$/As, corresponding to a charge carrier density
in a one-band model of 4.1 electrons/f.u..
The value of $R_H$ increases with decreasing temperature
to $-1.15\times10^{-10}$\ m$^3$/As at 80\,K.
If one assumes that there exists a residual magnetization increase in the high-field regime
its anomalous contribution is holelike. Therefore it will lead to an underestimation of $R_H$,
but {\it not} to a sign change. 
The observation of an {\it electronlike} ordinary Hall effect corresponding
to several electrons per f.u. in SFMO is the central result of this work. 

The anomalous Hall effect in ferromagnetic materials has two possible origins, an asymmetry of
scattering (skew scattering) or a sideward displacement of the center of weight of an
electron wave packet during the scattering process (side-jump), both due to spin-orbit interaction \cite{Berger70}.
The anomalous Hall effect is closely related to the longitudinal resistivity $\rho_{xx}$ by
\begin{equation}
	\label{scalelaw}
	R_A \mu_0 M = \gamma \rho_{xx}^n,
\end{equation}
but with a different exponent $n=1$ and $n=2$ for skew scattering and side jump, respectively.
The anomalous Hall coefficient can be extracted from the data by extrapolation the linear high-field data
to $B=0$ \cite{Camp+Fert}. The obtained value is then, according to Eq. \ref{RhoHall}, 
$R_A \mu_0 M_{Sat}:=\rho_{xy}^\star$. The resistivites $\rho_{xx}$ versus $\rho_{xy}^\star$ in a double
logarithmic plot show indeed in the case of SFMO a linear slope with $n=0.75$, indicating skew scattering.
Due to the opposite sign of the temperature coefficient of the resistivity between SFMO and the
manganites in the ferromagnetic regime, the anomalous Hall coefficient increases for SFMO with
decreasing temperature. In the manganites the anomalous Hall effect vanishes for very low temperatures,
because of increasing magnetic order, as expected by theory \cite{Smit55}. 
This is a further hint that in our SFMO thin films a full magnetic order is not obtained.

\section{ conclusion}
In summary we prepared high epitaxial thin films of the compound Sr$_{2}$FeMoO$_6$ 
with narrow rocking curves by pulsed laser deposition.
We performed detailed transport measurements of the diagonal and nondiagonal elements of the resistivity tensor from 
4\ K up to room temperature in magnetic fields up to 8\ T.
An electronlike ordinary Hall effect and a holelike anomalous Hall 
contribution were observed. These signs are reversed compared to the colossal magnetoresistive manganites. 
The value of the nominal charge carrier density at 300 K is four 
electrons per formula unit. A full magnetic order was not observed in our samples. 
 
\acknowledgments
We thank G.~Linker from Forschungszentrum Karlsruhe for the Rutherford backscattering analysis of the film stoichiometry.
This work was supported by the Deutsche Forschungsgemeinschaft through Project No. JA821/3.

\end{multicols}

\end{document}